\documentclass[11pt]{article}
\usepackage{amsmath,amssymb,color}

\def\ben{\begin{equation}}
\def\een{\end{equation}}

 \def\bd{\begin{document}} \def\ed{\end{document}}
\def\ds{\documentstyle} \let\fr=\frac \let\bl=\bigl \let\br=\bigr
\let\Br=\Bigr \let\Bl=\Bigl
\let\bm=\bibitem
\let\na=\nabla
\let\pa=\partial \let\ov=\overline
\newcommand{\be}{\begin{equation}}
\newcommand{\ee}{\end{equation}}
\def\ba{\begin{array}}
\def\ea{\end{array}}
\def\ft#1#2{{\textstyle{\frac{\scriptstyle #1}{\scriptstyle #2} } }}
\def\fft#1#2{{\frac{#1}{#2}}}
\def\del{\partial}
\def\vp{\varphi}
\def\sst#1{{\scriptscriptstyle #1}}
\def\oneone{\rlap 1\mkern4mu{\rm l}}
\def\td{\tilde}
\def\wtd{\widetilde}
\def\ie{{\it i.e.\ }}
\def\dalemb#1#2{{\vbox{\hrule height .#2pt
        \hbox{\vrule width.#2pt height#1pt \kern#1pt
                \vrule width.#2pt}
        \hrule height.#2pt}}}
\def\square{\mathord{\dalemb{6.8}{7}\hbox{\hskip1pt}}}
\def\i{{\rm i}}
\newcommand{\ho}[1]{$\, ^{#1}$}
\newcommand{\hoch}[1]{$\, ^{#1}$}
\newcommand{\bea}{\setlength\arraycolsep{2pt} \begin{eqnarray}}
\newcommand{\eea}{\end{eqnarray}}
\newcommand{\ra}{\rightarrow}
\newcommand{\lra}{\longrightarrow}
\newcommand{\Lra}{\Leftrightarrow}
\newcommand{\bp}{\tilde \beta^\prime}
\newcommand{\tr}{{\rm tr} }
\newcommand{\Tr}{{\rm Tr} }
\def\0{{\sst{(0)}}}
\def\1{{\sst{(1)}}}
\def\2{{\sst{(2)}}}
\def\3{{\sst{(3)}}}
\def\4{{\sst{(4)}}}
\def\5{{\sst{(5)}}}
\def\6{{\sst{(6)}}}
\def\7{{\sst{(7)}}}
\def\8{{\sst{(8)}}}
\def\m{{\sst{(m)}}}
\def\n{{\sst{(n)}}}
\def\cA{{{\cal A}}}
\def\cB{{{\cal B}}}
\def\cF{{{\cal F}}}
\def\cG{{{\cal G}}}
\def\cH{{{\cal H}}}
\def\tV{\widetilde V}
\def\tW{\widetilde W}
\def\tH{\widetilde H}
\def\tE{\widetilde E}
\def\tF{\widetilde F}
\def\tA{\widetilde A}
\def\im{{{\rm i}}}
\def\tY{{{\wtd Y}}}
\def\ep{{\epsilon}}
\def\vep{{\varepsilon}}
\def\bD{{{\bar D}}}
\def\R{{{\mathbb R}}}
\def\C{{{\mathbb C}}}
\def\H{{{\mathbb H}}}
\def\CP{{{\mathbb C}{\mathbb P}}}
\def\RP{{{\mathbb R}{\mathbb P}}}
\def\Z{{{\mathbb Z}}}
\def\bA{{{\mathbb A}}}
\def\bB{{{\mathbb B}}}
\def\bC{{{\mathbb C}}}
\def\bD{{{\mathbb D}}}
\def\bE{{{\mathbb E}}}
\def\bZ{{{\mathbb Z}}}
\def\Re{{{\frak{Re}}}}
\def\Im{{{\frak{Im}}}}
\def\cosec{{\,\hbox{cosec}\,}}
\def\Gm{{\Gamma_{\!\! -}}}
\def\Gp{{\Gamma_{\!\! +}}}
\def\stan{{standard }}
\def\nonstan{{supernumerary }}
\def\p{{\partial}}
\def\kdel#1{{\fft{\del}{\del#1}}}

\def\bog{{Bogomolny }}
\def\om{{\omega}}

\newcommand{\nnr}{\nonumber \\}
\newcommand{\pd}{\partial}
\newcommand{\ud}{\textrm{d}}
\newcommand{\dTH}{T^{\prime \, 0}_\textrm{H}}
\newcommand{\dOi}{\Omega^{\prime \, 0}_i}
\newcommand{\bx}{{\bf x}}

 \newcommand{\bpsi}{\bar\psi}
 \newcommand{\cL}{{\cal L}}
 \newcommand{\cR}{{\cal R}}
 \newcommand{\cT}{{\cal T}}
 \def\bra#1{\left\langle#1\right|}
 \def\ket#1{\left|#1\right\rangle}
 \def\brk#1{\left\langle#1\right\rangle}

\begin{document}

\begin{flushright}
\hfill{USTC-ICTS-10-07}

\end{flushright}

\vspace{25pt}
\begin{center}
{\large {\bf Instability by Chern-Simons and/or Transgressions}}

\vspace{15pt}

H. L\"u\hoch{\dagger\ddagger} and Zhao-Long Wang\hoch{\star}

\vspace{10pt}

\hoch{\dagger}{\it China Economics and Management Academy\\
Central University of Finance and Economics, Beijing, 100081}

\vspace{10pt}

\hoch{\ddagger}{\it Institute for Advanced Study, Shenzhen
University, Nanhai Ave 3688, Shenzhen 518060}

\vspace{10pt}

\hoch{\star}{\it Interdisciplinary Center for Theoretical Study,\\
University of Science and Technology of China, Hefei, Anhui 230026}

\vspace{40pt}

\underline{ABSTRACT}
\end{center}

It was demonstrated recently that there is an upper bound of the
Chern-Simons coupling of the five-dimensional Einstein-Maxwell
theory, beyond which the electrically charged AdS$_2\times S^3$
vacuum solution becomes unstable. We generalize the result to a
general class of gravity theories involving Chern-Simons and/or
transgression terms and find their upper bounds for stability. We
show that supergravities with AdS$\times$Sphere vacua satisfy the
bounds.

\vspace{15pt}

\thispagestyle{empty}





\newpage

\section{Introduction}

Chern-Simons and transgression terms associated with form fields are
common occurrences in supergravities.  Typically supergravities
allow all possible such terms but with the coupling strengths
dictated by the supersymmetry.  For example, the ``$\ft16$'' factor
of the Chern-Simons term in eleven-dimensional supergravity is
indeed fixed by the supersymmetry \cite{11dref}. It turns out this
term plays an important role in quantizing the supermembrane tension
\cite{dkl}.  It was demonstrated that the U-duality groups
$E_{n(+n)}$ of maximum supergravities coming from the $n$-torus
reduction would be broken to only the $GL(n,\R)$, had this
coefficient not been $\ft16$ \cite{cjlp}. This enhancement of global
symmetry from $GL(n,\R)$ to $E_{n(+n)}$ is crucial
\cite{clpspherered} for the consistent $S^7$ \cite{s7red1,s7red2} or
$S^4$ \cite{s4red1,s4red2} Kaluza-Klein reductions of
eleven-dimensional supergravity. The consistency requires a delicate
balance \cite{liupope} between the properties of the Killing vectors
in the spheres and the properties of eleven-dimensional
supergravity, including the ``$\ft16$'' factor.

  On the other hand, in most of the isotropic $p$-brane constructions
\cite{dkl,stainless} in string and M-theory, there is no
contribution from the Chern-Simons or the transgression terms. It is
intriguing to question whether such terms have any effects on the
$p$-brane physics.  In particular, we are interested in the
non-dilatonic $p$-branes whose decoupling limits give rise to
AdS$\times$Sphere backgrounds.  These solutions are expected, and in
some cases proven to be stable due to the supersymmetry they
preserve.  Nevertheless, it was recently observed \cite{nop} that
the Chern-Simons term could in principle provide a source of
instability. This was known in $D=3$ where topologically massive
gauge theory can developed a tachyon mode when the topological
Chern-Simons term is introduced \cite{tmg}. The example considered
in \cite{nop} was Einstein-Maxwell theory in five dimensions with a
generic Chern-Simons term. It was demonstrated that for the
electrically-charged AdS$_2\times S^3$ background, there is an upper
bound of the Chern-Simons coupling, beyond which tachyon modes
emerge. The coupling in supergravity satisfies this bound. For the magnetically-charged AdS$_3\times S^2$ solution,
there is no such instability.

    In this paper, we examine a large class of Chern-Simons and
transgression structures that could arise in supergravities.  We
relax the couplings to be arbitrary constants and discuss the
instability that could arise due to these terms.  In section 2, we
examine theories involving form fields with Chern-Simons and/or
transgression terms in flat spacetime backgrounds.  In general,
Chern-Simons terms always produce instability in backgrounds with
electric charges whilst the transgression terms produce instability
in magnetic backgrounds.  In section 3, we couple the system to the
Einstein-Hilbert action with a cosmological constant.  By making use
of the Breitenlohner-Freedman bound of AdS backgrounds, we derive
the the maximum coupling of the Chern-Simons and/or transgression
terms, beyond which instability will arise.  We apply the results in
various supergravities in section 4, and demonstrate that for
AdS$\times$Sphere backgrounds in supergravities, the bounds are
always satisfied; they would have been saturated had the momentum in
the internal direction be continuous. We conclude our paper in
section 5. In appendix A, we present a detailed linear analysis of
eleven-dimensional supergravity in AdS$_4\times S^7$ and
AdS$_7\times S^4$ backgrounds. We use this example to show that in
general the linear perturbation of the form fields that depends on
the Chern-Simons/transgression coupling decouples from the rest of
the perturbation modes including the graviton modes and hence can be
analyzed easily.  We give the condition for which these modes are no
longer decoupled from certain graviton modes.

\section{A general case in flat background}

\subsection{Either Chern-Simons or transgression term}

Let us consider a general case in flat spacetime background,
involving $(n,p,q)$-form field strengths.  The Lagrangian contains
only the kinetic terms and one Chern-Simons term, namely
\be {\cal L}_0 = - \ft12 {* H_{(n)}}\wedge H_{(n)} -  \ft12 {*
F_{(p)}}\wedge F_{(p)} - \ft12 {* G_{(q)}}\wedge G_{(q)} +
\alpha C_{(n-1)}\wedge F_{(p)}\wedge G_{(q)}\,.\label{flatlag1}\ee
where $H_{(n)}=dC_{(n-1)}$, $F_{(p)}=dA_{(p-1)}$,
$G_{(q)}=dB_{(q-1)}$.  It is clear that the spacetime dimension is
$D=n+p+q-1$.  The constant $\alpha$ measures the strength of the
Chern-Simons coupling.  It is well-known that the Chern-Simons and
transgression terms are sometimes related by the Hodge dualization.
This can certainly be done for (\ref{flatlag1}). To be specific, if
we perform the Hodge dual on the $H_{n}$ to become
$(D-n)$-form $\tilde H_{(D-n)}$, the Lagrangian becomes 
\be {\cal L}_0' = - \ft12 {* \tilde H_{(D-n)}}\wedge \tilde
H_{(D-n)} - \ft12 {* F_{(p)}}\wedge F_{(p)} - \ft12 {*
G_{(q)}}\wedge G_{(q)}\,.
\label{flatlag2}\ee 
In this Lagrangian there is no longer any  Chern-Simons term; however,
the $(D-n)$-form field strength is modified by a transgression term,
namely 
\be \tilde H_{(D-n)} = d\tilde C_{(D-n-1}) +\alpha\, A_{(p-1)}\wedge
G_{(q)}\,,\qquad dH_{(D-n)} =\alpha F_{(p)}\wedge G_{(p)}\,. \ee 
Thus there is no need for us to discuss the case with purely the
transgression term in detail, since it can be dualized to become a
Chern-Simons term.

    The equations of motion for (\ref{flatlag1}) are given by
\bea d{*H_{(n)}} &=& (-1)^{n D}\, \alpha\, F_{(p)}\wedge
G_{(q)}\,,\cr d{*F_{(p)}} &=& (-1)^{pq}\, \alpha\, H_{(n)} \wedge
G_{(q)}\,,\cr d{*G_{(q)}} &=& \alpha\, H_{(n)}\wedge F_{(p)}\,.
\eea 
In terms of index notation, the equations are given by 
\bea \partial_{I} H^{IJ_1\cdots J_{n-1}} &=& \fft{\alpha}{p! q!}
\varepsilon^{J_1\cdots J_{n-1} K_1\cdots K_p L_1\cdots K_q}
F_{K_1\cdots K_p}G_{L_1\cdots L_q}\,,\cr
\partial_I F^{IJ_1\cdots J_{p-1}} &=& \fft{(-1)^{p(D-q)} \alpha}{n! q!}
\varepsilon^{J_1\cdots J_{p-1} K_1\cdots K_n L_1\cdots K_q}
H_{K_1\cdots K_n}G_{L_1\cdots L_q}\,,\cr
\partial_I G^{IJ_1\cdots J_{q-1}} &=& \fft{(-1)^{qD} \alpha}{n! p!}
\varepsilon^{J_1\cdots J_{q-1} K_1\cdots K_n L_1\cdots K_p}
H_{K_1\cdots K_n}G_{L_1\cdots L_p}\,. \eea
Here the tensor $\varepsilon$ is a pure number in flat background
and we adopt the convention that $\varepsilon_{012\cdots}=1$.

Let us consider a background with vanishing form fields $F_{(p)}$
and $F_{(q)}$ but non-zero $H_{(n-1)}$. The equations for the linear
fluctuation $(U,V)$ for $(F,G)$ are then given by
\bea
\partial_I U^{IJ_1\cdots J_{p-1}} &=& \fft{(-1)^{p(D-q)} \alpha}{n! q!}
\varepsilon^{J_1\cdots J_{p-1} K_1\cdots K_n L_1\cdots K_q}
H_{K_1\cdots K_n}V_{L_1\cdots L_q}\,,\cr
\partial_I V^{IJ_1\cdots J_{q-1}} &=& \fft{(-1)^{qD} \alpha}{n! p!}
\varepsilon^{J_1\cdots J_{q-1} K_1\cdots K_n L_1\cdots K_p}
H_{K_1\cdots K_n}U_{L_1\cdots L_p}\,,\label{genlinear1} \eea
In this section, we examine the possible instability due to the
Chern-Simons term for constant electric or magnetic $H_{(n)}$, or
dyonic in $D=2n$ dimensions.  In the last subsection, we shall
consider a system with both Chern-Simons and transgression terms. It
should be emphasized that for our discussion it is equivalent to
turn on each one of the $(n,p,q)$ forms.  In special cases where two
or all three field strengths are the same, some combinatoric factors
can be altered without changing the essential conclusion.

\subsection{Electric $H_{(n)}$}

     We first consider the case with $H_{(n)}$ being electric,
namely
\be H_{(n)} = (-1)^n E dt\wedge dx^1\wedge\ldots\wedge
dx^{n-1}\,,\ee 
where $E$ is a constant.  The minus factor in the above plays no
essential role and it is merely to make the intermediate formulae
better looking.  The whole spacetime is split into $n$-dimensional
sub-spacetime $T$ with coordinates $x^\mu$ and $(d=D-n=p+q-1)$
dimensional space $S$ with coordinates $y^i$. The equations of
motion (\ref{genlinear1}) become
\bea
\partial_I U^{Ii_1\cdots i_{p-1}} &=& -\fft{\alpha E}{q!}
\varepsilon^{i_1\cdots i_{p-1} j_1\cdots j_q} V_{j_1\cdots
j_q}\,,\cr
\partial_I V^{I i_1\cdots i_{q-1}} &=& -\fft{(-1)^{pq}
\alpha E}{p!}\varepsilon^{i_1\cdots i_{q-1} j_1\cdots j_p}
U_{j_1\cdots j_p}\,,\cr
\partial_I U^{I J_1\cdots J_{p-2} \mu}&=&0\,,\qquad
\partial_I V^{I J_1\cdots J_{q-2} \mu}=0\,.\label{UVeom}
\eea
Note that $U$ and $V$ satisfy the following Bianchi identity
\be
\partial_{[I}U_{J_1\cdots J_p]} =0 =\partial_{[I}
V_{J_1\cdots J_q]}\,. \ee

Let us define 
\be \widetilde U^{i_1\cdots i_{q-1}} = \fft{1}{p!}
\varepsilon^{i_1\cdots i_{q-1}}{}_{j_1\cdots j_p} U^{j_1\cdots
j_p}\,,\qquad \widetilde V^{i_1\cdots i_{p-1}} = \fft{1}{q!}
\varepsilon^{i_1\cdots i_{p-1}}{}_{j_1\cdots j_q} V^{j_1\cdots
j_q}\,. \label{UVsdef}\ee
This implies that 
\be U^{i_1\cdots i_{p}} = \fft{(-1)^{p(q-1)}}{(q-1)!}
\varepsilon^{i_1\cdots i_{p}}{}_{j_1\cdots j_{q-1}} \widetilde
U^{j_1\cdots j_{q-1}}\,,\quad V^{i_1\cdots i_{q}} =
\fft{(-1)^{q(p-1)}}{(p-1)!} \varepsilon^{i_1\cdots
i_{q}}{}_{j_1\cdots j_{p-1}} \widetilde V^{j_1\cdots j_{p-1}}\,.
\ee 
Acting on the first and second equations in (\ref{UVeom}) by
$\varepsilon_{\ell i_1\cdots i_{p-1} k_1\cdots
k_{q-1}}\partial^\ell$ and $\varepsilon_{\ell i_1\cdots i_{q-1}
k_1\cdots k_{p-1}}\partial^\ell$ respectively, we have
\bea &&\square\widetilde U^{j_1\cdots j_{q-1}} + \fft{\alpha
E}{(p-1)!} \varepsilon^{j_1\cdots j_{q-1}}{}_{\ell i_1\cdots
i_{p-1}} \partial_\ell \widetilde V^{i_1\cdots i_{p-1}}\,,\cr
&&\square\widetilde V^{j_1\cdots j_{p-1}} + \fft{(-1)^{pq} \alpha
E}{(q-1)!} \varepsilon^{j_1\cdots j_{p-1}}{}_{\ell i_1\cdots
i_{q-1}}\partial_\ell \widetilde U^{i_1\cdots i_{q-1}}\,, \eea
where
\be \square = \del^\mu\del_\mu + \del^\ell\del_\ell\,. \ee
The above two equations can be expressed in terms of form language,
namely 
\be \square \widetilde U_s + \alpha E\,  {*_s d\widetilde V_s}
=0\,,\qquad \square \widetilde V_s + (-1)^{pq} \alpha E\, {*_s
d\widetilde U_s}=0\,.\label{UVformexpE}
\ee 
The subscript ``$s$'' denotes that the forms and the Hodge dual are
defined in the $S$-space. In the momentum basis, $e^{ip_\mu x^\mu +
i k_i y^i}$, we have \be
\begin{pmatrix} (M^2-k^2)I_{N_p} & \alpha E J_{p,q}\cr
                \alpha E J_{p,q}^\dagger & (M^2 - k^2) I_{N_q}
                \end{pmatrix}
\begin{pmatrix} \widetilde U_s \cr \widetilde V_s \end{pmatrix}=0
\label{matrixE} \ee
where $M^2=-p^\mu p_\mu$, $k^2=k^i k_i$, $N_p=C_d^{p-1}$ and
$N_q=C_d^{q-1}$. Also $J$ is the $N_p\times N_q$ matrix of momenta
$k_i$ and $I_N$ is the $N\times N$ identity matrix.

Let us look at some specific examples of $J$. For convenience we
may arrange $(\widetilde U,\widetilde V)$ in the lexical order.  For
$q=1$, $J_{p,1}$ is a row vector of dimension $p$ and the components
are given by 
\be(J_{p,1})_i={\rm i} (-1)^{i+1} k_{p+1-i}\,.\ee 
The next simplest example is $p=2$ and $q=2$, for which we have 
\be J_{2,2}={\rm i} \begin{pmatrix} 0 & -k_3 & k_2\cr k_3 & 0 &
-k_1\cr -k_2& k_1 & 0 \end{pmatrix} \ee 

    The matrix in (\ref{matrixE}) is hermitian and hence guaranteed to
have real eigenvalues.  The mass of possible tachyon modes can be
determined by the vanishing of the determinant of the matrix, which
leads to the condition 
\be M^2 - k^2 \pm \alpha E k=0\,.\ee 
(The $M=0$ solution with non-vanishing $k$ is incompatible with the
Bianchi identity.) Thus there are tachyon modes for $0<k<\ft12
|\alpha E|$. In section 3, we shall analyze the system coupled to
gravity, in which case the constant electric $F_{(n)}$ can support
an AdS$_n\times S^{D-n}$ background.

\subsection{Magnetic $H_{(n)}$}

We now examine the case with $H_{(n)}$ being magnetic, namely
\be H_{(n)} = (-1)^{nD+1} B dy^1\wedge\ldots\wedge dy^n\,. \ee 
We split the whole spacetime into two parts: the $(d=p+q-1)$
dimensional spacetime $T$, with coordinates $x^\mu$ and the
$n$-dimensional space $S$ with coordinates $y^i$. The equations for
the linear perturbations (\ref{genlinear1}) become 
\bea &&\partial_I U^{I\mu_1\cdots \mu_{p-1}} + \fft{\alpha B}{q!}
\varepsilon^{\mu_1\cdots\mu_{p-1}}{}_{\nu_1\cdots\nu_q}
V^{\nu_1\cdots \nu_q}=0\,,\cr 
&&\partial_I V^{I\mu_1\cdots \mu_{q-1}}+ \fft{(-1)^{pq}\alpha B}{p!}
\varepsilon^{\mu_1\cdots\mu_{q-1}}{}_{\nu_1\cdots\nu_p}
U^{\nu_1\cdots \nu_p}=0\,,\cr 
&&\partial_I U^{I J_1\cdots J_{p-2} i}=0\,,\qquad
\partial_I V^{I J_1\cdots J_{q-2} i} =0\,.
\eea 
We define 
\be \widetilde U^{\mu_1\cdots \mu_{q-1}} =
\fft{1}{p!}\varepsilon^{\mu_1\cdots\mu_{q-1}}{}_{\nu_1\cdots\nu_p}
U^{\nu_1\cdots \nu_p}\,,\quad 
\widetilde V^{\mu_1\cdots \mu_{p-1}} =
\fft{1}{q!}\varepsilon^{\mu_1\cdots\mu_{p-1}}{}_{\nu_1\cdots\nu_q}
U^{\nu_1\cdots \nu_q}\,. \ee 
This implies that 
\bea U^{\mu_1\cdots\mu_p} &=&-\fft{(-1)^{p(q-1)}}{(q-1)!}
\varepsilon^{\mu_1\cdots\mu_p}{}_{\nu_1\cdots\nu_{q-1}} \widetilde
U^{\nu_1\cdots\nu_{q-1}}\,,\cr V^{\mu_1\cdots\mu_q}
&=&-\fft{(-1)^{q(p-1)}}{(p-1)!}
\varepsilon^{\mu_1\cdots\mu_q}{}_{\nu_1\cdots\nu_{p-1}} \widetilde
V^{\nu_1\cdots\nu_{p-1}}\,. \eea 
The equations for $\tilde U$ and $\tilde V$ can be cast into the
same form as (\ref{UVformexpE}), except that now Hodge dual and the
forms are defined within the $T$-spacetime, namely
\be \square \widetilde U_t + \alpha E\,  {*_t d\widetilde V_t}
=0\,,\qquad \square \widetilde V_t + (-1)^{pq} \alpha E\, {*_t
d\widetilde U_t}=0\,.\label{UVformexpM}
\ee 
Here the subscript $t$ labels the $T$-spacetime.  In this case, the
characteristic equation for the mass $M$ and momentum modular $k$ is
given by 
\be M^2 - k^2 + \alpha B M =0\,.\ee 
It is thus clear that there is no tachyon mode.

\subsection{Dyonic $H_{(n)}$}

When $D=2n$, the field strength $H_{(n)}$ can be both electric and
magnetic, namely 
\be H_{(n)} = (-1)^n E\, dt\wedge dx^1\wedge\ldots \wedge dx^{n-1} -
B dy^1\wedge\ldots \wedge dy^n\,. \ee 
The $D$-dimensional spacetime is split into the $n$-dimensional
spacetime $T$ with coordinates $x^\mu$ and the $n$-dimensional space $S$
with coordinates $y^i$.  It is straightforward to derive the
linearized equations of motion, which contain the following
\bea \square \widetilde U_s + \alpha E\,  {*_s d\widetilde V_s}
=0\,,&& \square \widetilde V_s + (-1)^{pq} \alpha E\, {*_s
d\widetilde U_s}=0\,,\cr 
\square \widetilde U_t + \alpha B\, {*_t d\widetilde V_t} =0\,,&&
\square \widetilde V_t + (-1)^{pq} \alpha B\, {*_t d\widetilde
U_t}=0\,.\label{UVformexpEM} \eea 
Thus there are tachyon modes associated with $\widetilde U_s$ and
$\widetilde V_s$.

\subsection{Both Chern-Simons and transgression terms}

There can be both Chern-Simons and transgression terms associated
with the same field strength $H_{(n)}$.  The corresponding equations
of motion and Bianchi identity are characterized by 
\be d{*H_{(n)}} =(-1)^{nD} \alpha F_{(p)}\wedge G_{(q)}\,,\qquad
dH_{(n)}= \beta \tilde F_{(\tilde p)}\wedge \tilde G_{(\tilde q)}
\,.\ee
Let $\widetilde X$ and $\widetilde Y$ be associated with $\tilde F$ and
$\tilde G$ in the same way as $\widetilde U$ and $\widetilde V$
associated with $F$ and $G$.  When the $H_{(n)}$ is electric, we
have
\bea \square \widetilde U_s + \alpha E\,  {*_s d\widetilde V_s}
=0\,,&& \square \widetilde V_s + (-1)^{pq} \alpha E\, {*_s
d\widetilde U_s}=0\,,\cr 
\square \widetilde X_t + \beta E\, {*_t d\widetilde Y_t} =0\,,&&
\square \widetilde Y_t + (-1)^{pq} \beta E\, {*_t d\widetilde
X_t}=0\,. \eea 
When the $H_{(n)}$ is magnetic, we have 
\bea \square \widetilde U_t + \alpha B\,  {*_t d\widetilde V_t}
=0\,,&& \square \widetilde V_t + (-1)^{pq} \alpha B\, {*_t
d\widetilde U_t}=0\,,\cr 
\square \widetilde X_s + \beta B\, {*_s d\widetilde Y_s} =0\,,&&
\square \widetilde Y_s + (-1)^{pq} \beta B\, {*_s d\widetilde
X_s}=0\,.\eea 
Thus in this case, there are tachyon modes regardless whether
$H_{(n)}$ is electric or magnetic.

\section{Coupled to Gravity}

In the previous section we consider the instability arising from the
Chern-Simons or transgression terms of form fields in the flat
Minkowskian background.  We now examine the effect of gravity
coupled to this system.  We first consider the Lagrangian
\be {\cal L}=(R-2\Lambda) {*\oneone} + {\cal L}_0\,. \ee
where ${\cal L}_0$ takes the same form as (\ref{flatlag1}). The
first term in the above is the Einstein-Hilbert term with a
cosmological constant $\Lambda$.  Let us first consider the AdS$\times
S^{D-n}$ vacuum solution supported by the electric $H_{(n)}$; it is
given by
\bea ds_D^2 &=& a^2 ds_n^2 + b^2 d\Omega_{D-n}^2\,,\qquad H_{(n)} =
E\, a^n \epsilon_{(n)}\,,\cr 
E^2&=&\fft{2(n-1)}{a^2} + \fft{2(D-n-1)}{b^2}\,,\qquad -2\Lambda =
\fft{(n-1)^2}{a^2} - \fft{(D-n-1)^2}{b^2}\,, \label{adssphereE}
\eea 
where $ds_n^2$ and $d\Omega_{(D-2)}^2$ are the unit AdS$_n$  and
$S^{D-n}$ metrics, satisfying $R_{\mu\nu} = -(n-1) g_{\mu\nu}$ and
$R_{ij} = (D-n-1) g_{ij}$ respectively.

As discussed in the end of appendix A, in the special case for $n=2$
with $H_{(2)}=F_{(2)}$ and/or $H_{(2)}=G_{(2)}$, the gravitational
fluctuation couples with that of the 2-form field strengths through
the Chern-Simons coupling $\alpha$. This case was studied in
\cite{nop} for five-dimensional Einstein-Maxwell theory. In more
generic cases, as we show in appendix A, the gravitational
perturbation is independent of $\alpha$. Since the purpose of the
paper is to examine the effect of $\alpha$ on the stability of the
AdS$\times$Sphere solutions, there is no need for us to present the
perturbation of the metric here. The relevant modes are the same as
the one discussed in the flat background, namely $\widetilde U_s$
and $\widetilde V_s$ defined by (\ref{UVsdef}). (Here the subscript
$s$ denotes that the quantities carry only the indices in the
$S^{D-n}$ directions.) They now satisfy
\be \Delta \widetilde U_s + \alpha E\,  {*_s d\widetilde V_s}
=0\,,\qquad \Delta \widetilde V_s + (-1)^{pq} \alpha E\, {*_s
d\widetilde U_s}=0\,,\label{UVformexpE1}
\ee
where $\Delta=-(dd^\dagger + d^\dagger d)$ is the Laplace operator with
respect to the AdS$\times$Sphere background. This implies that the mass
of the possible tachyon modes is again determined by
\be M^2 - k^2 \pm \alpha E k =0\,. \ee 
Thus the minimum value of the mass for the tachyon modes is given by
\be M_{\rm min} = - \ft14 \alpha^2 E^2\,. \ee 
For this to satisfy the Breitenlohner-Freedman (BF) bound of the
AdS$_n$ spacetime, namely
\be M_{\rm min}^2 \ge M^2_{\rm BF} = - \fft{(n-1)^2}{4 a^2}\,, \ee 
we have 
\be \alpha^2 \Big(1 + \fft{2a^2\Lambda}{(n-1)(D-2)}\Big) \le
\fft{(n-1)(D-n-1)}{2(D-2)}\,. \ee 
There are two cases arising. The first case is when $\Lambda\le 0$,
for which the $\alpha$ has a maximum value, namely 
\be \alpha^2\le \alpha^2_{\rm max} \equiv \fft{(n-1)(D-n-1)}{2(D-2)}
\,.\label{alphamax1} \ee 
Once this condition is satisfied, there is no instability due to the
Chern-Simons term for all the allowed parameter regions of the
AdS$\times$Sphere solutions in (\ref{adssphereE}). The second case
is when $\Lambda>0$.  In additional to the condition that $\alpha$ has
to be smaller than $\alpha_{\rm max}$, there is
a further requirement that the AdS radius
has to be sufficiently small.  For a given $\alpha <\alpha_{\rm
max}$, the maximum radius for the AdS$_n$ is given by 
\be a_{\rm max}^2 = \fft {(n-1)(D-2)}{2\Lambda}
\Big(\fft{\alpha^2_{\rm max}}{\alpha^2} -1\Big)\,. \ee 
Solutions with $a>a_{\rm max}$ suffers from the instability due to
the Chern-Simons term.

      For the magnetic AdS$_{D-n}\times S^n$ solution, it is
straightforward to show that the mass formula is then given by 
\be M^2 - k^2 \pm \alpha B M=0\,,\quad\rightarrow\quad M^2=
\Big(\sqrt{k^2 +\ft14 \alpha^2 B^2} \pm \ft14 \alpha B\Big)^2\,.
\ee 
Thus there is no instability due to the Chern-Simons term.

 It is clear that if the $H_{(n)}$ has only the transgression
term instead of the Chern-Simons term, the electric solution will
always be stable whilst the magnetic solution will be stable only if
the analogous condition discussed above with $E$ replaced by
magnetic flux parameter $B$ is satisfied. When $H_{(n)}$ has both
Chern-Simons and transgression terms, the above conditions have to
be satisfied in order to avoid instability regardless whether the
$H_{(n)}$ is electric or magnetic.

If $H_{(n)}$ is self-dual, then the corresponding AdS$_n\times S^n$
is given by
\bea ds_{2n}^2 &=& a^2 ds_n^2 + b^2 d\Omega_n^2\,,\qquad F_{(n)} = E
(a^n \epsilon_{(n)} + b^n \Omega_{(n)})\,,\cr E^2 &=&
(n-1)\Big(\fft{1}{a^2} - \fft{1}{b^2}\Big)\,,\qquad -2\Lambda =
(n-1)^2\Big(\fft{1}{a^2} - \fft{1}{b^2}\Big)\,. \eea 
Thus, for $\Lambda \le 0$, there is no instability due to the
Chern-Simons and/or transgression terms as long as we have $\alpha
\le \alpha_{\rm max}$ with
\be \alpha_{\rm max}^2 = \ft12 (n-1)\,,\label{alphamax2} \ee
Note that this expression for $\alpha_{\rm max}^2$ has a factor 2
difference compared to that in (\ref{alphamax1}) specializing in
$D=2n$. For $\Lambda >0$, when this condition is satisfied,
there can still be instable AdS$_n\times S^n$ solutions as long as
the AdS radius is larger than $a_{\rm max}$, where 
\be a_{\rm max}^2=\fft{(n-1)^2}{\Lambda} \Big(\fft{\alpha_{\rm
max}^2}{\alpha^2} - 1\Big)\,. \ee 

\section{Applications in supergravities}

We now apply the results obtained in the previous section in
supergravities. Let us first examine eleven-dimensional
supergravity, which has AdS$_4\times S^7$ and AdS$_7\times S^4$
vacuum solutions. The detailed analysis of linearized perturbation
in these backgrounds were presented in appendix A.
Eleven-dimensional supergravity has the Chern-Simons term, given by
\be {\cal L}_{FFA} = \alpha A_\3\wedge F_\4\wedge F_\4\,, \ee 
where $|\alpha|=1/6$.  The situation is slightly
different from the examples discussed in sections 2 and 3, where the
$(n,p,q)$ forms are all different. The equation of motion for $F_\4$
now produces a factor 3, namely 
\be d{*F_4} = 3\alpha F_\4 \wedge F_\4\,. \ee 
Furthermore, if we consider $F_\4=\bar F_\4 + f_\4$, where $\bar
F_\4$ is the background and $f_\4$ is a small perturbation, the
above equation picks another factor 2, {\it i.e.} 
\be d\delta({*\bar F_\4}) + d{\bar *f_\4} =
6\alpha \bar F_4\wedge f_\4\,.
\ee
As shown in appendix A, the first term plays no role, and it
follows that the stability condition (\ref{alphamax1}) is modified
by a factor 6 and becomes
\be |\alpha|\le |\alpha_{\rm max}| =\ft16\,. \ee 
Thus the bound of Chern-Simons coupling for the stable AdS$_4\times
S^7$ is saturated naively by eleven-dimensional supergravity. (See
Appendix for further discussion.) For the magnetic AdS$_7\times
S^4$, there is no such a bound.

Type IIB supergravity has an AdS$_5\times S^5$ vacuum solution
supported by the self-dual 5-form field strength $H_\5$. It couples
to the R-R and NS-NS 3-form field strengths by both Chern-Simons and
transgression terms, leading to the equations of motion
\be d{*H_\5} = - F_\3^1 \wedge F_\3^2\,,\qquad d{*F_\3^1} =
H_\5\wedge F^2_\3\,,\qquad d{*F_\3^2} = - H_\5\wedge F_\3^1\,, \ee
where $F_\3^1=dA_\2^1$, $F_\3^2=dA_\2^2$ are the R-R and NS-NS
3-forms respectively. There is a subtlety with the 5-form
normalization; it enters the energy-momentum tensor with a
$1/\sqrt2$ factor, leading to $1/2$ of the contribution of the usual
convention of the  tensor.  This implies that the
condition of stability (\ref{alphamax2}) is modified by a factor 2,
leading to
\be |\alpha|\le |\alpha_{\rm max}|=1\,. \ee 
Thus the bound is also satisfied by type IIB supergravity.

     Another example is the AdS$_3\times S^3$ vacuum solution of
six-dimensional supergravities supported by a self-dual 3-form. In
maximum supergravity, there can be such Chern-Simons terms in the
form of $A_\2\wedge F_\2^1\wedge F_\2^2$ and/or analogous
transgression terms.  In less than maximum supergravities, they
become $\ft12 A_\2\wedge F_\2\wedge F_\2$.  At the level of
equations of motion for the 2-forms, the coupling constant for
Chern-Simons and/or transgression terms can be viewed as $\alpha=1$
in both cases. Since the stability bound for the AdS$_3\times S^3$
supported by the self-dual 3-form is given by (\ref{alphamax2}), it
follows that the bound is also satisfied by six-dimensional
supergravities.

     The stability condition for the electric AdS$_2\times S^3$ was
studied in \cite{nop}.  If one would naively apply the condition
(\ref{alphamax1}) taking into account that the Chern-Simons term
involves the same $U(1)$ vector field, one would obtain the bound
$\alpha_{\rm max} =1/(6\sqrt3)$, which is half of the supergravity
value. However, it turns out that in this special case, as explained
in appendix A, the linear perturbation of the the Maxwell field
cannot be decoupled from all the linear perturbation modes of the
metric. The mass formula is thus modified by the inclusion of
certain relevant graviton modes. The consequence is that the bound
is also satisfied by five-dimensional supergravity \cite{nop}.

\section{Conclusions}

An important issue in the AdS/CFT correspondence is the stability of the
AdS$\times$Sphere vacuum solutions in supergravities.  These solutions
are expected to be stable by the argument of supersymmetry.  However
the abundant Chern-Simons and/or transgression terms in supergravities
could in principle produce a source of instability, even though they play
no role in the construction of these solutions.  In this paper we relax
the couplings of the Chern-Simons/transgression terms to be arbitrary
constants and show that there indeed are upper bounds for these couplings
beyond which instability occurs.

    We show that in general the tachyon modes arise from the linear
perturbation of the form fields and decouple from the gravitational
modes. However, in some special cases tachyon modes can also involve
the gravitational modes and the analysis can be more involved.  The
conclusion is that the couplings of Chern-Simons/transgression terms
in supergravities examined all satisfy their stability bounds. It
would have saturated the bounds had the momenta in the internal
direction be continuous. The result clearly indicates the special
feature of supergravities and suggests that the AdS/CFT
correspondence may only be valid within a sound theory such as
supergravities.

      Our focus of analysis has been the non-dilatonic $p$-branes
whose decoupling limits give rise to AdS$\times$Sphere backgrounds.
It is of interest to investigate the analogous stability condition
for dilatonic $p$-branes whose decoupling limit give rise to a
product of a domain wall spacetime and a sphere, which may shed
light on the domain wall/QFT correspondence.

\section*{Acknowledgement}

We would like to thank the organizer and the participants of the advanced
workshop ``Dark Energy and Fundamental Theory", supported by the Special
Fund for Theoretical Physics from the National Natural Science Foundation
of China for useful discussions.

\appendix
\section{Instability analysis for $D=11$ supergravity}

The Lagrangian for the bosonic sector of eleven -dimensional
supergravity is given by
\be {\cal L} = R{*\oneone} -\ft12 *F_\4\wedge F_\4 + \ft16
A_\3\wedge F_\4\wedge F_\4\,, \ee 
where $F_\4=dA_\3$.  We shall consider a more general Lagrangian by
adding a cosmological constant $\Lambda$ and replacing the
Chern-Simons coupling $\ft16$ by an arbitrary constant $\alpha$. The
resulting equations of motion are modified, given by
\begin{eqnarray}
R_{IJ}-\ft29\Lambda g_{IJ} &=& \ft1{12} F^2_{IJ}
-\ft{1}{144} F^2\, g_{IJ}\,,\cr 
\partial_{I}(\sqrt{-g} F^{IJ_1J_2J_3})&=&\sqrt {-g}
\,\nabla_{I} F^{IJ_1J_2J_3} = \fft{3\alpha}{(4!)^2}
\ep^{J_1J_2J_3K_1\dots K_8}\, F_{K_1\dots K_4}\,
F_{K_5\dots K_8}\,.
\end{eqnarray}
where $\ep^{0,1\dots,10}=-1$.  The system admits AdS$_4\times S^7$
and AdS$_7\times S^4$ vacuum solutions.  We shall analyze the
stability conditions for both vacua.

\subsection{AdS$_4\times S^7$}

The AdS$_4\times S^7$ solution is given by
\begin{eqnarray}
ds^2 &=& \bar g_{\mu\nu}
dx^{\mu}dx^{\nu}=\fft{R_4^2}{r^2} ds_4^2 + R_7^2d\Omega_7^2 \,,\qquad
F_\4=E\,R_4^4 \epsilon_\4\,,\cr
E &=& \sqrt{\fft6{R_4^2} +\fft{12}{R_7^2}}\,,\qquad \Lambda =
-\fft9{2R_4^2} + \fft{18}{R_7^2}\,,
\end{eqnarray}
where $\epsilon_\4$ is the volume form for the $ds_4^2$.  The metrics
$ds_4^2$ and $d\Omega_7^2$ describe the unit AdS$_4$ and $S^7$
respectively. In other words, the curvatures of the background are
given by
\begin{eqnarray}
\bar R_{\mu\nu\rho\sigma}&=&-\fft1{R_4^2}\left(\bar g_{\mu\rho}\bar
g_{\nu\sigma}-\bar g_{\nu\rho}\bar g_{\mu\sigma}\right)\,,\qquad
\bar R_{ijkl}=\fft1{R_7^2}\left(\bar g_{ik}\bar g_{jl}-\bar g_{jk}
\bar g_{il}\right)\,,\cr
\bar R_{\mu\nu}&=&-\fft3{R_4^2}\bar g_{\mu\nu}\,,\quad\bar R_{ij}
=\fft6{R_7^2}\bar g_{ij}\,,\quad\bar R=-\fft{12}{R_4^2}+\fft{42}{R_7^2}
\,.
\end{eqnarray}
It is clear that we have split the whole spacetime index $I,J=0,1,\dots,
10$ into $\mu,\nu,\dots=0,1,2,3$ to label the indices
in AdS$_4$ and $i,j,\dots=4,5,\dots,10$ to label the indices in the
$S^7$ directions. The fluctuations of the metric and the 4-form are
denoted by
\begin{eqnarray}
g_{IJ}&=&\bar g_{IJ}+h_{IJ}\,,\qquad
F_{IJKL}=\bar F_{IJKL} +f_{IJKL}\,.\label{pert}
\end{eqnarray}
We find
\begin{eqnarray}
R_{IJ}&=&\bar R_{IJ}+
\ft12\left(\bar\nabla_{K}\bar\nabla_{I}h_{J}^{~\,K}+\bar\nabla_{K}
\bar\nabla_{J}h_{I}^{~\,K} -\bar\nabla^{K}\bar\nabla_{K} h_{IJ}-
\bar\nabla_{I}\bar\nabla_{J}h^{K}_{~\,K}\right)\,,\cr
\delta( \sqrt{-g})&=&\ft12\sqrt{-\bar g}h^{I}_{~\,I}\,,
\end{eqnarray}
Our purpose is to discuss the instability due to the
Chern-Simons $FFA$ term, hence we first consider the
linearized equation of motion for the gauge field.
It will become apparent presently that it is advantageous
to adopt the traceless gauge
\begin{equation}
h\equiv
h^{I}{}_{I}=0\,.
\end{equation}
It follows that $\delta(\sqrt{-g})=0$.
The variation of the 4-form field strength around the background
is given by
\begin{equation}
\delta F^{IJKL}= f^{IJKL}-\bar F_{M}{}^{JKL}h^{MI}-\bar
F^{I}{}_M{}^{KL}h^{MJ} -\bar F^{IJ}{}_M{}^{L}h^{MK}
-\bar F^{IJK}{}_Mh^{ML}\,.
\end{equation}
Specifically, we have
\begin{equation}
\delta F^{\mu\nu\rho\sigma} = f^{\mu\nu\rho\sigma} -
E\bar\varepsilon^{\mu\nu\rho\sigma}h^{\lambda}_{\,~\lambda}\,,\quad
\delta F^{\mu\nu\rho i} = f^{abci} - E\bar\varepsilon^{\mu\nu\rho\sigma}
h_{\sigma}{}^i\,,\quad
\delta F^{IJij}=f^{IJij}\,.
\end{equation}
The linearized equations for the gauge field perturbations are
\begin{eqnarray}
&&\partial_{\sigma}\left(\sqrt{-\bar g}(f^{\sigma\mu\nu\rho}-
E\bar\varepsilon^{\sigma\mu\nu\rho} h^\lambda_{\,~\lambda})\right)
+\partial_{i}\left(\sqrt{-\bar
g}(f^{i\mu\nu\rho}-E\bar\varepsilon^{\sigma\mu\nu\rho}h_{\sigma}^{~\,i})
\right)=0\,, \cr &&\partial_{\rho}\left(\sqrt{-\bar g}
(f^{\rho\mu\nu i}- E\bar\varepsilon^{\rho\mu\nu
\sigma}h_{\sigma}^{~\,i})\right) +\partial_{j}\left(\sqrt{-\bar g}
f^{j\mu\nu i}\right)=0\,,\cr &&
\partial_{\nu}\left(\sqrt{-\bar g} f^{\nu\mu ij}\right)
+\partial_{k}\left(\sqrt{-\bar g} f^{k\mu ij}\right)=0\,,\cr
&&\label{eqfi}\partial_{\mu}\left(\sqrt{-\bar g} f^{\mu ijk}\right)
+\partial_{l}\left(\sqrt{-\bar g} f^{lijk}\right)
= -\fft{\alpha}4
E\fft{R_4^4}{r^4}\ep^{ijkl_1l_2l_3l_4}f_{l_1l_2l_3l_4}\,.
\end{eqnarray}
Note that the parameter $\alpha$ appears only in the last equation
in (\ref{eqfi}), which can also be expressed as
\begin{eqnarray}
\label{eqfi1} \bar\nabla^{\mu}f_{\mu ijk} +\bar\nabla^{l}f_{lijk}
+\fft{\alpha}4
E\bar\varepsilon_{ijk}^{\,~\,~\,~l_1l_2l_3l_4}f_{l_1l_2l_3l_4}=0\,.
\end{eqnarray}
Acting $\bar\varepsilon_{m_1m_2m_3}{}^{mijk}\bar\nabla_m$ on
(\ref{eqfi1}), we find
\begin{eqnarray}
\label{eqfi2}
0\!&=&\!\fft1{3!}\bar\varepsilon_{m_1m_2m_3}{}^{mijk}\bar\nabla_m
(\bar\nabla^{\mu }f_{\mu ijk} +\bar\nabla^{l}f_{lijk} +\fft{\alpha}4
E\bar\varepsilon_{ijk}^{\,~\,~\,~l_1l_2l_3l_4}f_{l_1l_2l_3l_4})
\cr\!&=&\!\fft1{3!}\bar\varepsilon_{m_1m_2m_3}{}^{mijk}
\left[\bar\nabla^{\mu }\bar\nabla_mf_{\mu ijk}
+\bar\nabla^{l}\bar\nabla_mf_{lijk}+\bar
R^{l~~n}_{~\,m\,~l}f_{nijk}+ \bar R^{l~~n}_{~\,m\,~i}f_{lnjk}\right.\cr
\cr&&~~~~~~~~~~~~~~~~~~~~\left.+\bar R^{l~~n}_{~\,m\,~j}f_{link}+\bar
R^{l~~n}_{~\,m\,~k}f_{lijn}+\fft{\alpha}4
E\bar\nabla_m(\bar\varepsilon_{ijk}^{\,~\,~\,~l_1l_2l_3l_4}
f_{l_1l_2l_3l_4})\right]
\cr\!&=&\!\fft1{3!}\bar\varepsilon_{m_1m_2m_3}{}^{mijk}
\left[\fft1{4}\bar\nabla^{\mu }\bar\nabla_{\mu }f_{mijk}
+\fft1{4}\bar\nabla^{l}\bar\nabla_lf_{mijk}-\fft3{R^2_7}f_{mijk}\right.\cr
   &&~~~~~~~~~~~~~~~~~~~~\left. +\fft{\alpha}4
E\bar\nabla_m(\bar\varepsilon_{ijk}^{\,~\,~\,~l_1l_2l_3l_4}
f_{l_1l_2l_3l_4})\right]
\cr\!&=&\!(\Box_4 +\triangle_7)f_{m_1m_2m_3} +{\alpha}
E\bar\varepsilon_{m_1m_2m_3}{}^{mijk}\bar\nabla_mf_{ijk} \,,
\end{eqnarray}
where we have defined
\begin{eqnarray}
f^{m_1m_2m_3}=\ft1{4!}\bar\varepsilon^{m_1m_2m_3mijk}f_{mijk}\,.
\end{eqnarray}
We also have used the Bianchi identity as well as the explicit
form of the Laplace operator $\triangle_7=-(d_7d_7^\dagger+
d_7^\dagger d_7)$ acting on a tensor:
\begin{eqnarray}
&&\nabla_{[I} f_{I_1I_2I_3I_4]}=\partial_{[I} f_{I_1I_2I_3I_4]}=0\,,
\cr
&&\triangle_7f_{ijk}=\bar\nabla^{m}\bar\nabla_mf_{ijk}-\bar
R^m_{~~i}f_{mjk}-\bar R^m_{~~j}f_{imk}-\bar R^m_{~~k}f_{ijm} +2\bar
R^{m\,~n}_{~~i~~j}f_{mnk}\cr
&&~~~~~~+2\bar R^{m\,~n}_{~~i~~k}f_{mjn} +2\bar
R^{m\,~n}_{~~j~~k}f_{imn}=\bar\nabla^{m}
\bar\nabla_mf_{ijk}-\fft{12}{R^2_7}f_{ijk}\,.
\end{eqnarray}

The terms appearing in linear perturbation of the energy-momentum
tensor are
\begin{eqnarray}
&&\delta F^2_{\mu\nu} = \fft12 \bar g_{\mu\nu}
\bar\varepsilon^{\rho_1\rho_2\rho_3\rho_4}f_{\rho_1\rho_2\rho_3\rho_4} +
6E^2(\bar g_{\mu\nu} h^{\rho}{}_{\rho} - h_{\mu\nu})\,,\quad
\delta F^2_{\mu i} = E \bar \varepsilon_{\mu}{}^{\nu\rho\sigma} f_{i
\nu\rho\sigma}\,,\quad \delta F^2_{ij}=0\,,\cr
&&\delta(\ft1{4!}F^2g_{IJ}) = \bar g_{IJ} \fft{2E}{4!}
\bar\varepsilon^{\rho_1\rho_2\rho_3\rho_4}
f_{\rho_1\rho_2\rho_3\rho_4}+E^2(\bar g_{IJ} h^{\rho}{}_{\rho}
- h_{IJ})\,.
\end{eqnarray}
Note that only the perturbations $f_{\rho_1\rho_2\rho_3\rho_4}$
and $ f_{i\nu\rho\sigma}$ contribute to the  tensor
and the parameter $\alpha$ does not appear in the Einstein
equations. Combining with (\ref{eqfi}), we conclude that $f_{ijk}$
are decoupled from other fluctuations in the linear order,
with only the $f_{ijk}$ affected by the parameter $\alpha$.
Therefore, the mass of the possible tachyon modes is determined by
\begin{eqnarray}\label{masssform}
M^2-k^2\pm6\alpha Ek=0
\end{eqnarray}
It is clear that there are tachyon modes for $0<k<|3\alpha E|$. The
most negative tachyon mode happens at $k=|3\alpha E|$, where
$M^2=-(3\alpha E)^2$. The Breitenlohner-Freedman bound of
AdS$_{d+1}$ is
\begin{eqnarray} m^2_{\rm BF}=-\fft{d^2}{4R_4^2}\,. \end{eqnarray}
To avoid the physical instability, we need $(3\alpha E)^2=9\alpha^2
\left(\fft6{R_4^2}+\fft{12}{R_7^2}\right)\leq\fft{9}{4R_4^2}$. For
eleven-dimensional supergravity we have $\alpha=\fft1{6}$, and hence
the bound would be saturated if the internal momentum $k$ had been
continuous.  However, to saturate the bound requires that
$k=|3\alpha E|$ be a possible mode in the spectrum of $f_{ijk}$ on
the $S^7$. The spectrum of $f_{ijk}$ on $S^7$ is given by
$k=\pm{(l+6)}/{R_7}$, where $l=0,1,2,\dots$. On the other hand,
given $\alpha=\fft1{6}$ and $\Lambda=0$, we find that $3\alpha
E=3/R_7$. Therefore, although the Breitenlohner-Freedman bound is
saturated by the naive minimum of the mass formula (\ref{masssform})
for eleven-dimensional supergravity, it could not be saturated by
any modes of $f_{ijk}$ on $S^7$. The lowest mode for eleven-dimensional supergravity
corresponds to $M=0$ which occurs when $k=\pm{6}/{R_7}$. In fact, to make sure the lowest mode from the spectrum of $f_{ijk}$ on the $S^7$ satisfying the Breitenlohner-Freedman bound, we only need $\alpha\leq5/24$.

      It is worth emphasizing that although the traceless gauge
is normally a ``wasteful'' choice, it serves our purpose in that
the potential tachyon modes due to the Chern-Simons term decouple
manifestly. The disadvantage is that the
discussion for the linearized Einstein equations become more complicated,
for which it is convenient to take the De Donder gauge.  Since the
the graviton modes play no role in our conclusion of the stability,
we shall not present them here.

\subsection{AdS$_7\times S^4$}

The AdS$_7\times S^4$ solution supported by the magnetic $F_\4$ is
given by
\begin{eqnarray}
ds^2&=&\bar g_{IJ}dx^{I}dx^{J}=R_7^2 ds_7^2 + R_4^2d\Omega_4^2
\,,\qquad \bar F= {B\,R_4^4}\,\Omega_\4\,,\cr
B&=&\sqrt{\fft6{R_4^2}+\fft{12}{R_7^2}}\,,\qquad
\Lambda=\fft9{2R_4^2}-\fft{18}{R_7^2}\,.
\end{eqnarray}
Correspondingly we have
\begin{eqnarray}
\bar R_{\mu\nu\rho\sigma}&=&-\fft1{R_7^2}\left(\bar g_{\mu\rho}\bar
g_{\nu\sigma}-\bar g_{\nu\rho}\bar g_{\mu\sigma}\right)\,,\qquad\bar
R_{ijkl}=\fft1{R_4^2}\left(\bar g_{ik}\bar g_{jl}-\bar g_{jk}\bar
g_{il}\right)\,,\cr \bar R_{\mu\nu} &=&-\fft6{R_7^2}\bar
g_{\mu\nu}\,,\quad\bar R_{ij}=\fft3{R_4^2}\bar g_{ij}\,,\quad\bar
R=\fft{42}{R_4^2}-\fft{12}{R_7^2}\,,
\end{eqnarray}
where we have split the whole spacetime index $I,J=0,1,\dots,
10$ into $\mu,\nu,\dots=0,1,\dots,6$ to label the indices
in AdS$_7$ and $i,j,\dots=7,\dots,10$ to label the indices in the
$S^4$ directions. The fluctuations of the metric and the
4-form are again given by (\ref{pert}). Under the traceless
gauge we have $\delta( \sqrt{-g})= 0\,.$
The variation involving the 4-form is given by
\begin{eqnarray}
&&\delta F^{\mu\nu IJ}=f^{\mu\nu IJ}\,,\qquad
\delta F^{\mu ijk}=f^{\mu ijk}-B\bar\varepsilon_{l}^{~\,ijk} \,h^{l\mu}\,,
\cr
&&\delta F^{ijkl}= F_{~\,~\,~m}^{ijk}h^{ml}=f^{ijkl}-B\bar
\varepsilon^{ijkl} h_m^{\,~m}\,,\cr
&&\delta F^2_{\mu\nu} =0\,,\qquad
\delta F^2_{\mu i}=B\bar\varepsilon_{i}^{\,~jkl}f_{\mu jkl}\,,\qquad
\delta F^2_{ij}=\ft12B\,\bar g_{ij}\bar
\varepsilon^{nklm}f_{nklm}-6B^2(\bar g_{ij}h_k^{\,~k}-h_{ij})
\,,\cr
&&\delta (\fft1{4!}F^2g_{IJ}) =\fft{2B}{4!}\bar g_{IJ}\bar
\varepsilon^{i_1i_2i_3i_4}f_{i_1i_2i_3i_4}- B^2\left(\bar g_{IJ}
h^i_{\,~i}-h_{IJ}\right)\,.
\end{eqnarray}
The linearized equations for the 4-form perturbation are
\begin{eqnarray}
&&\partial_{\lambda}\left(\sqrt{-\bar g}f^{\lambda\mu\nu\rho}\right)
+\partial_{i}\left(\sqrt{-\bar g}
f^{i\mu\nu\rho}\right)=\fft{\alpha}4 B\sqrt{\bar
g_4}\,\epsilon^{\mu\nu\rho
\lambda_1\lambda_2\lambda_3\lambda_4}
f_{\lambda_1\lambda_2\lambda_3\lambda_4}\,,\cr
&&\partial_{\rho}\left(\sqrt{-\bar g} f^{\rho\mu\nu i}\right)
+\partial_{j}\left(\sqrt{-\bar g} f^{j\mu\nu i}\right)=0\,,\cr
&&\partial_{\nu}\left(\sqrt{-\bar g} f^{\nu\mu ij}\right)
+\partial_{k}\left(\sqrt{-\bar g} f^{k\mu
ij}+B\bar\varepsilon_{l}^{~\,kij} \,h^{l\mu }\right)=0\,,\cr
&&\label{eqbi}\partial_{\mu}\left(\sqrt{-\bar g}( f^{\mu
ijk}-B\bar\varepsilon_{l}^{~\,ijk}\,h^{l\mu})\right)
+\partial_{l}\left(\sqrt{-\bar g} (f^{lijk}-B\bar\varepsilon^{lijk}
h_m^{\,~m})\right)=0\,.
\end{eqnarray}
The first equation in (\ref{eqbi}) is the only one that depends on
the parameter $\alpha$; it can be expressed as
\begin{eqnarray}
\label{eqbi1} \bar\nabla^{i}f_{i\mu\nu\rho} +\bar\nabla^{\sigma}
f_{\sigma\mu\nu\rho} -\fft{\alpha}4
B\bar\varepsilon_{\mu\nu\rho}^{\,~\,~\,~\sigma_1\sigma_2\sigma_3\sigma_4}
f_{\sigma_1\sigma_2\sigma_3\sigma_4}=0\,.
\end{eqnarray}
Acting $\bar\varepsilon_{\sigma_1\sigma_2\sigma_3\sigma_4}^{~~~~~~~~~~
\sigma\mu\nu\rho}\bar\nabla_d$ on (\ref{eqbi1}), we find
\begin{eqnarray}
\label{eqbi2}
(\Box_4 +\triangle_7)f_{\sigma_1\sigma_2\sigma_3} -{\alpha}
B\bar\varepsilon_{\sigma_1\sigma_2\sigma_3}^{~~~~~~~~~\sigma\mu\nu\rho}
\bar\nabla_{\sigma}f_{\mu\nu\rho}=0
\,,
\end{eqnarray}
where we have defined
\begin{eqnarray}
f^{\sigma_1\sigma_2\sigma_3}=\ft1{4!}\bar\varepsilon^{\sigma_1
\sigma_2\sigma_3\sigma\mu\nu\rho}f_{\sigma\mu\nu\rho}\,
\end{eqnarray}
and we have also used the Bianchi identity as well as the explicit
form of the Laplace operator $\triangle_7=-(d_7d_7^\dagger+
d_7^\dagger d_7)$ acting on a tensor
\begin{eqnarray}
\triangle_7f_{\mu\nu\rho}=\bar\nabla^{\sigma}\bar\nabla_{\sigma}
f_{\mu\nu\rho}+\fft{12}{R^2_7}f_{\mu\nu\rho}\,.
\end{eqnarray}
Analog with the case for electric flux, only the perturbation
$f_{i_1i_2i_3i_4}$ and ${f_{\mu jkl}}$ contribute
to the energy-momentum tensor and the parameter $\alpha$ does not appear in
the Einstein equations. Therefore, $f_{\mu\nu\rho}$ are
decoupled with other fluctuations in linear order. The mass of the
these modes is decided by
\begin{eqnarray}
M^2-k^2\pm6\alpha B M=0\Rightarrow
M^2=\left(\sqrt{k^2+(3\alpha B)^2}\pm3\alpha B\right)^2\,.
\end{eqnarray}
There is no tachyon instability.

 An important lesson we learn from the above linear analysis is that
the potential tachyon modes $f_{ijk}$ due to the Chern-Simons term
decouple at the linear order from the rest of the perturbations.
This is obvious for the generic $(n,p,q)$-system discussed in
sections 2 and 3, where only $H_{(n)}$ is non-vanishing and
$F_{(p)}$ and $G_{(q)}$ differ from $H_{(n)}$.  In this example,
however, the $(n,p,q)$-forms are all the same as one, namely the
$F\4$. The background $\bar F_\4$ is non-vanishing and the 4-form
perturbation does couple with the gravitational perturbation.
Nevertheless, as can be seen from (\ref{eqfi}) and (\ref{eqbi}), for
an $n$-form, its equation of motion has $(n-1)$ free indices, but
only the modes involving at least $(n-2)$ free indices in the
parallel directions of the background flux couple with the graviton
modes. These modes are independent of the parameter $\alpha$ for
$n>2$. Thus the potential tachyon modes decouple from the rest modes
and satisfy a simple equation (\ref{UVformexpE1}) for $n>2$.  It is
clear that the above argument breaks down for $n=2$ and the tachyon
modes can no longer decouple from some gravitational modes. This
situation happens in the Einstein-Maxwell Chern-Simons theory in
five dimensions. The equations of motion for the tachyon modes are
more complicated and the result was announced in \cite{nop}. It
turns out that there is again an upper bound of the Chern-Simons
coupling for stability and the naive minimum of the mass formula in
five-dimensional supergravity would have saturated the bound, if the
momentum in the internal direction had been continuous. However, to
saturate the bound requires that $k=|1/r_3|$ while the corresponding
spectrum on the $S^3$ is given by $k=|(l+2)/r_3|$ with $l=0,1,2\dots
$. Therefore, the bound could not be saturated by any modes on the
sphere and the real lowest mode corresponds to $M=0$. In fact, it is
a common feature in all the supergravities we have examined.

\end{document}